\documentclass[]{spie}  

\usepackage{amsmath,amsfonts,amssymb}
\usepackage{graphicx}
\usepackage[colorlinks=true, allcolors=blue]{hyperref}

\title{Upgrading SPHERE with the second stage AO system SAXO+: non-common path aberrations estimation and correction}

\author[a]{Johan Mazoyer}
\author[a]{Charles Goulas}
\author[a]{Fabrice Vidal} 
\author[b]{Isaac Bernardino Dinis}
\author[c]{Julien Milli} 
\author[d]{Michel Tallon}
\author[a]{Raphaël Galicher} 
\author[e]{Oliver Absil}
\author[d]{Clémentine Béchet}
\author[a]{Anthony Boccaletti} 
\author[a]{Florian Ferreira}
\author[d]{Maud Langlois}
\author[f]{Patrice Martinez}
\author[g]{Laurent Mugnier} 
\author[f]{Mamadou N’diaye}
\author[e]{Gilles Orban de Xivry} 
\author[h]{Axel Potier}
\author[d]{Isabelle Tallon-Bosc}
\author[i]{Arthur Vigan}

\affil[a]{ LESIA, Observatoire de Paris, Universit\'e PSL, CNRS, Sorbonne Universit\'e, Universit\'e de Paris, Meudon, France} 
\affil[b]{ Départment d’astronomie de l’Université de Genève, 51 ch. des Maillettes Sauverny, 1290 Versoix, Switzerland} 
\affil[c]{ Université Grenoble Alpes, CNRS, IPAG, 38000, Grenoble, France} 
\affil[d]{ Space Sciences, Technologies \& Astrophysics Research (STAR) Institute, Université de Liège, Allée du Six Août 19c, B-4000 Liège, Belgium} 
\affil[e]{ Univ Lyon, Univ Lyon1, Ens de Lyon, CNRS, Centre de Recherche Astrophysique de Lyon UMR5574, F-69230, Saint-Genis-Laval, France} 
\affil[f]{ Université Côte d’Azur, Observatoire de la Côte d’Azur, CNRS, Laboratoire Lagrange, France} 
\affil[g]{ DOTA, ONERA, Université Paris Saclay, BP 72, 92322 Châtillon cedex, France} 
\affil[h]{ Division of Space and Planetary Sciences, University of Bern, Gesellschaftsstrasse 6, 3012 Bern, Switzerland} 
\affil[i]{ Aix Marseille Univ, CNRS, CNES, LAM, Marseille, France}

\authorinfo{Send correspondence to johan.mazoyer@obspm.fr}

\begin{document} 
\maketitle

\begin{abstract}
SAXO+ is a planned enhancement of the existing SAXO, the VLT/ SPHERE adaptive optics system, deployed on ESO’s Very Large Telescope. This upgrade is designed to significantly enhance the instrument's capacity to detect and analyze young Jupiter-like planets. The pivotal addition in SAXO+ is a second-stage adaptive optics system featuring a dedicated near-infrared pyramid wavefront sensor and a second deformable mirror. This secondary stage is strategically integrated to address any residual wavefront errors persisting after the initial correction performed by the current primary AO loop, SAXO. However, several recent studies clearly showed that in good conditions, even in the current system SAXO, non-common path aberrations (NCPAs) are the limiting factor of the final normalized intensity in focal plane, which is the final metric for ground-based high-contrast instruments. This is likely to be even more so the case with the new AO system, with which the AO residuals will be minimized. Several techniques have already been extensively tested on SPHERE in internal source and/or on-sky and will be presented in this paper. However, the use of a new type of sensor for the second stage, a pyramid wavefront sensor, will likely complicate the correction of these aberrations. Using an end-to-end AO simulation tool, we conducted simulations to gauge the effect of measured SPHERE NCPAs in the coronagraphic image on the second loop system and their correction using focal plane wavefront sensing systems. We finally analyzed how the chosen position of SAXO+ in the beam will impact the evolution of the NCPAs in the new instrument.  
 
\end{abstract}


\section{INTRODUCTION: THE SAXO+ INSTRUMENT}
\label{sec:intro}

The primary objective of direct imaging is to capture images of circumstellar environments, including exoplanets, debris disks, and protoplanetary disks. High contrast imaging technique allows precise determination of the position of objects around stars and measures the light they reflect or emit. For exoplanets, direct imaging facilitates the determination of orbital parameters and the physicochemical properties of their atmospheres. 

However, this method is challenging due to the vast difference in luminosity (ranging from $10^{-4}$ to $10^{-10}$) and the small angular separations (ranging from a few hundredths of an arcsecond to a few arcseconds for the nearest stars) between a planet and its host star. Consequently, fewer than 1\% of the approximately 5,000 exoplanets discovered to date have been directly imaged. The last generation of high contrast instruments (SPHERE \cite{beuzit2019_SPHEREExoplanetImager}, GPI\cite{macintosh2014_FirstLightGemini}, SCEXAO\cite{lozi2020_StatusSCExAOInstrument} and MagAO-x\cite{Males2018SPIE_MagAO-X}) have been used to detect planetary mass companion around close-by stars. These instrument combine extreme adaptive optics (AO) systems, coronagraphic devices, and sophisticated dedicated calibration and post-processing methods, to remove starlight and detect circumstellar objects down to a few resolutions elements. 

As part of SPHERE's guaranteed time observations, the SpHere INfrared survey of Exoplanets (SHINE) observed around 600 stars. Initial SHINE analysis of the first 150 targets indicates that giant planets are rare beyond 10 AU \cite{vigan2021_SPHEREInfraredSurvey}. Conversely, radial velocity surveys suggest that most planets are located within 3 to 10 AU \cite{Lagrange2023AA_distributiongiants}. However, the current capabilities of SPHERE \cite{langlois2021_SPHEREInfraredSurvey} in this separation range are still insufficient to image planets identified through other techniques such as astrometry, radial velocity or microlensing. Therefore, further enhancements in contrast at short angular separations are essential to provide a comprehensive understanding of planet distribution across all separations. Another limitation of SPHERE is the maximum magnitude allowed by SAXO, SPHERE's AO system \cite{fusco2014_FinalPerformanceLessonlearned}. The current limit is an G magnitude of 14, with moderate Strehl Ratio (SR), which prevents the detection of fainter and redder targets, host of planet-forming disks observed in the submillimeter. 

\begin{figure} [ht]
\begin{center}
\begin{tabular}{c} 
\includegraphics[width=0.8\linewidth]{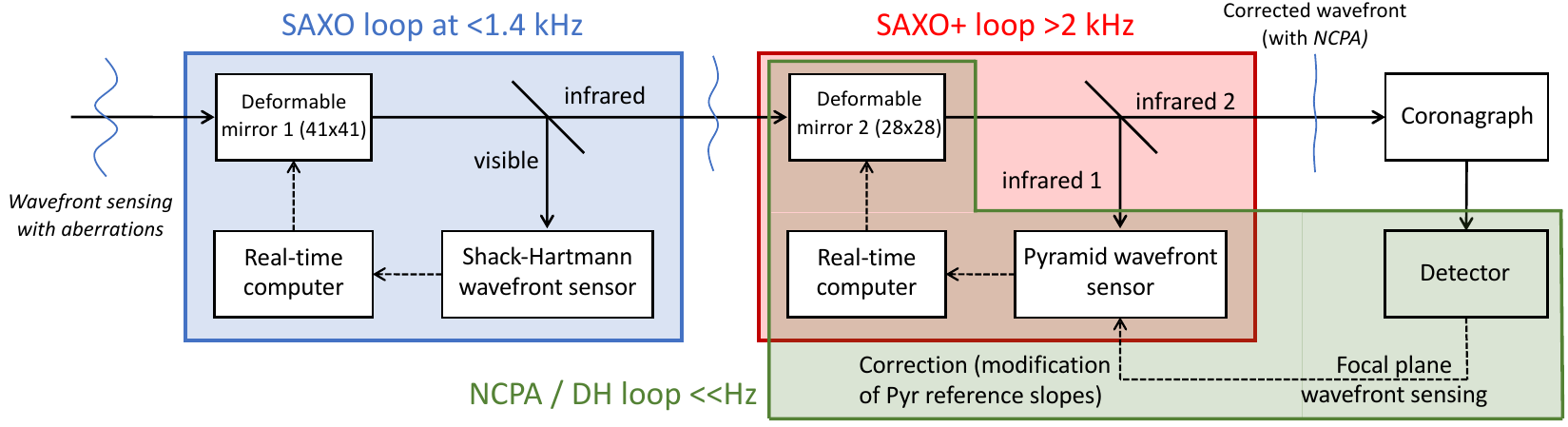}
\end{tabular}
\end{center}
\caption[]{\label{fig:saxo_3loops} \textbf{SAXO+ outline}. In blue: current SAXO system. In red: second loop of SAXO+. In green: NCPA correction (adapted from Goulas et al. 2022\cite{goulas2023_SAXOUpgradeSecond}).}
   \end{figure} 

To circumvent these limitations, an upgrade of SAXO has been proposed: SAXO+\cite{boccaletti2022_UpgradingHighContrast}. The main upgrade is the addition of a second AO stage. SAXO+ is also necessary as an on-sky technological demonstrator for the Planetary Camera and Spectrograph (PCS) roadmap for the Extremely Large Telescope\cite{kasper2021_PCSRoadmapExoearth}. To detect light reflected by planets, potentially as small as Super Earths, PCS will require an efficient extreme AO system. Currently, achieving contrast as low as $10^{-9}$ at 0.1 arcsecond with PCS will require the development of a complex two-stage AO instrument.

Figure~\ref{fig:saxo_3loops} shows the outline of SAXO+. SAXO is shown in blue box, with a Shack–Hartmann (SH) Wavefront Sensor (WFS) in visible light and a loop around 1KHz. In the red box is shown the proposed second loop using a Pyramid WFS (PyWFS). To gain contrast at short separations around bright stars, the 2nd stage will run faster, with a typical range of 2 to 3 kHz. The exact frequencies of the 2 loops, depending on the star brightness and observing conditions, are currently optimized using simulations.  

Non Common Path Aberrations (NCPA), aberrations introduced between the sensing and the science channels, degrade the performance of all AO systems. By creating quasi-static speckles, in the science focal plane, they are the limiting factor of the performance of high contrast imagers in good atmospheric conditions. This is likely to be even more so the case with SAXO+, with which the AO residuals will be minimized. 

NCPA compensation is a standard procedure for all AO systems. It usually involves introducing an offset in the wavefront sensor (WFS) signals, specifically in the SH slope reference measurements, to counteract the aberrations. This method is highly effective when the WFS measurements remain linear within the range of the NCPA amplitude (like for SH WFS). However, for a Pyramid WFS, this linear range is quite restricted and even if the NCPA are perfectly estimated, the accuracy of a given offset of the normalized pixels on PyWFS detector (referred to as PyWFS response hereafter) to correct for them decrease with increasing NCPA levels, and therefore part of the NCPAs are not corrected (or in extreme cases it can even degrade the correction). The estimation of the optical gains of the NCPA modes can be used to extend this range \cite{korkiakoski2008_ImprovingPerformancePyramid,deo2019_TelescopereadyApproachModala,esposito2020_OnskyCorrectionNoncommon,chambouleyron2020_PyramidWavefrontSensor}. 

In the next section, we address the problem of estimation of NCPA with coronagraphic instruments and the recent successful measurement obtained of the last 5 years with SPHERE. These methods have allowed us to precisely constrain and characterize the SPHERE NCPA and their evolution, and we will present these characteristic in Section~\ref{sec:measurement_evol}. Finally, in Section~\ref{sec:ncpa_corr}, we will present preliminary results of the correction of realistic NCPA aberrations in simulation.

\section{FOCAL PLANE WAVEFRONT SENSING ON SPHERE}

Wavefront aberration estimation of NCPA down to the final focal plane is a complicated problem. The fact that we cannot use a specific channel to measure the aberrations (if one doesn't want to add more NCPAs) most classical WFS techniques (SH, pyramid WFS) cannot be used. These methods are classically addressed using a range of techniques using directly the final science detector (i.e. IRDIS in this case) to measure NCPAs, often using so-called phase-diversity methods \cite{mugnier2006_PhaseDiversityTechnique}, which do not require additional hardware. 

\begin{figure} [ht]
\begin{center}
\begin{tabular}{c} 
\includegraphics[width=0.7\linewidth]{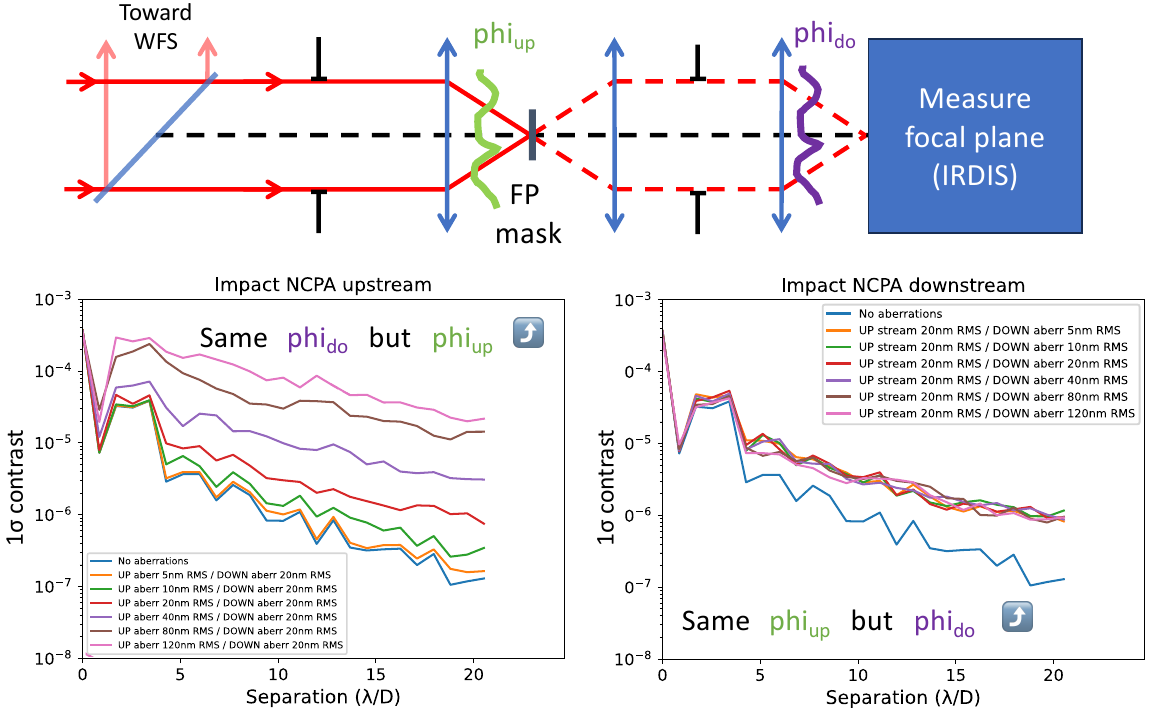}
\end{tabular}
\end{center}
\caption[]{\label{fig:abere_upanddown} \textbf{Upstream and downstream NCPAs}. \textbf{Top:} Schematic of a coronagraph showing the upstream NCPAs (green, introduced between the WFS channel beam splitter and the focal plane mask) and the downstream NCPAs (purple, introduced between the focal plane mask and the detector). \textbf{Bottom: }contrast curves of simulated IRDIS instrument to show impact of NCPAs (assuming perfect residuals phase after the beam splitter). Left: impact of increasing upstream NCPAs level with constant downstream NCPAs. Right: impact of increasing downstream NCPAs with constant upstream NCPAs. These simulations were performed using the \texttt{Asterix} coronagraphic simulation package\cite{mazoyer_AsterixSimulator}.}
\end{figure} 

The problem is even more complicated for coronagraphic instruments, as different types of NCPA are encountered: upstream of the coronagraphic mask or downstream of the coronagraphic mask. Figure~\ref{fig:abere_upanddown} shows the position of these NCPAs in the system (top) and simulations revealing their relative impacts on the final contrast in the focal plane (bottom). These clearly show that upstream NCPAs are quickly degrading performance, while downstream NCPAs impact on contrast is often negligible (for ground-based instruments). This means that a naive approach by removing the coronagraph and measure the full NCPAs using IRDIS will not provide acceptable results, as such a method would estimate the sum of upstream and downstream NCPAs and will not minimize the amount of NCPAs seen by the coronagraphic mask. 

Several methods have been proposed to estimate NCPAs in coronagraphic instruments (see reviews \cite{groff2016_MethodsLimitationsFocal, galicher2024_ImagingExoplanetsCoronagraphic}), either in the coronagraphic mask FP (only upstream NCPAs are measured and corrected and downstream NCPAs are ignored) or directly in the coronagraphic image (upstream and downstream NCPAs are separately measured and corrected using their relative impact on the final image). Some of these techniques have already been extensively tested on SPHERE in internal source or on-sky and will be presented in the following sections. As SAXO+ is a technological demonstration for PCS, the goal is to try to test and compare most or all of these techniques in the context of a 2 loop systems with a pyramid. 

\subsection{Initial phase diversity algorithm}
\label{subsec:init_sauvagecorrec}
Initial calibration of NCPA on SPHERE were performed using an optimized phase diversity algorithm described in Sauvage et al. (2011) \cite{sauvage2011_SPHERENoncommonPath}, which was integrated in ESO initial calibration of the instrument. The principle was to first introduce a phase diversity with SAXO DM to measure all aberrations (upstream and downstream) using IRDIS. Then, a second phase diversity estimation of the downstream aberrations only was performed. This measurement used a calibration source positioned at the coronagraphic mask focal plane, and the defocus was produced by moving the source along the optical axis and uses IRDIS images, the upstream aberrations were then deduced by subtraction of those measurements. 

This method is still to this date the baseline for SPHERE estimation of NCPA. In the 2014 final performance paper of SAXO\cite{fusco2014_FinalPerformanceLessonlearned}, it is indicated that after compensation of NCPA with this method, “measured SR is larger than 99 \% in H band, which means less than 20 nm RMS of residuals and less than 5 nm on the 50 first modes.” 

In addition to the complexity of this method, its main default is that classical phase diversity can only evaluate properly low order aberrations: the dynamic range necessary to image faint speckles created by high-order aberrations without saturating the PSF at the center exceed the one possible by IRDIS. Even before the commissioning, another method of coronagraphic phase diversity was proposed. 

\subsection{Coronagraphic phase diversity}
\label{subsec:coffee}
\begin{figure} [ht]
\begin{center}
\begin{tabular}{c  c} 
\includegraphics[width=0.75\linewidth]{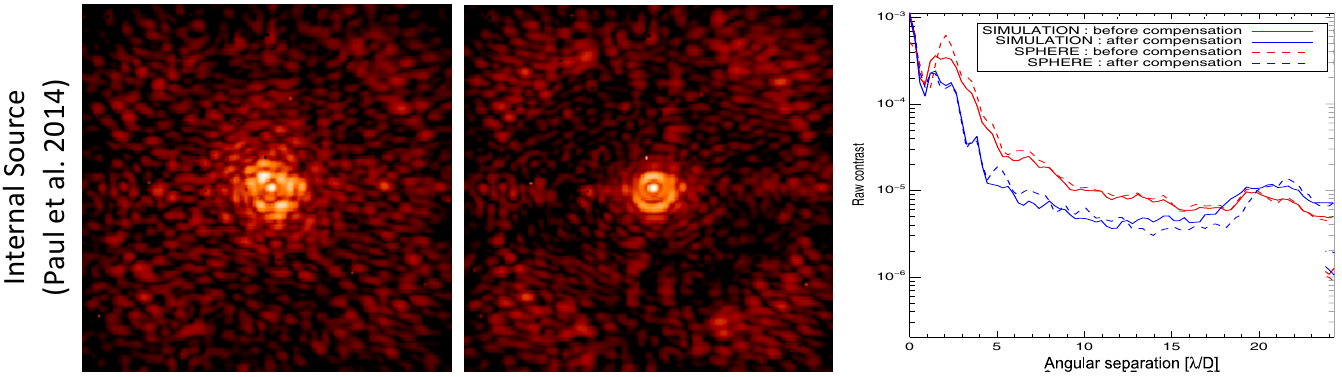}
\end{tabular}
\end{center}
\caption[]{\label{fig:coffee_sphere_paul14} \textbf{Coronagraphic phase diversity NCPA correction on SPHERE internal} source during integration in Grenoble, France, before the shipping to Paranal. From left to right: before compensation, after compensation, and contrast curves comparison (azimuthal mean profiles of the image normalized to the off-axis PSF). Figure adapted from Paul et al. (2014)\cite{Paul14_COFFEE_SPHERE}.}
\end{figure} 
The COronagraphic Focal plane waveFront Estimation for Exoplanet detection (COFFEE, Sauvage et al. (2012)\cite{sauvage2012_CoronagraphicPhaseDiversity}). COFFEE method is a coronagraphic phase diversity estimation of the aberrations both upstream and downstream of the coronagraph using two coronagraphic focal plane images (usually a defocused images followed by a normal image). The phase is then reconstructed using a Bayesian approach relying on a coronagraphic imaging model. It was tested in 2014 during integration in Grenoble (France), before the shipping to Paranal, on internal source by Paul et al. (2014)\cite{Paul14_COFFEE_SPHERE} with successful reduction of the NCPA as shown on Figure~\ref{fig:coffee_sphere_paul14}. It was never tested on sky, but simulations and testbed validation were performed to update the model in the presence of AO residuals \cite{herscovici2019_coffee_atmo}. 

Its main advantages compared to the baseline method described in the previous section is that it does not require any modification of the instrument hardware and is a much less complex procedure. Additionally, coronagraphic phase diversity allows the estimation of  aberrations of higher order than classical phase diversity \cite{paul2013_CoronagraphicPhaseDiversity}.

However, to introduce defocus on the mirror during AO correction, one has to modify the reference responses of the WFS. But as discussed in the previous section, introducing offsets on the PyWFS reference responses is not as straightforward as for SH WFS and there is still uncertainty if the defocus can be introduced on the DM this way and if yes, at which precision. 

\subsection{Zernike Wavefront sensing}
\label{subsec:zelda}
\begin{figure} [ht]
\begin{center}
\begin{tabular}{c  c} 
\includegraphics[width=0.75\linewidth]{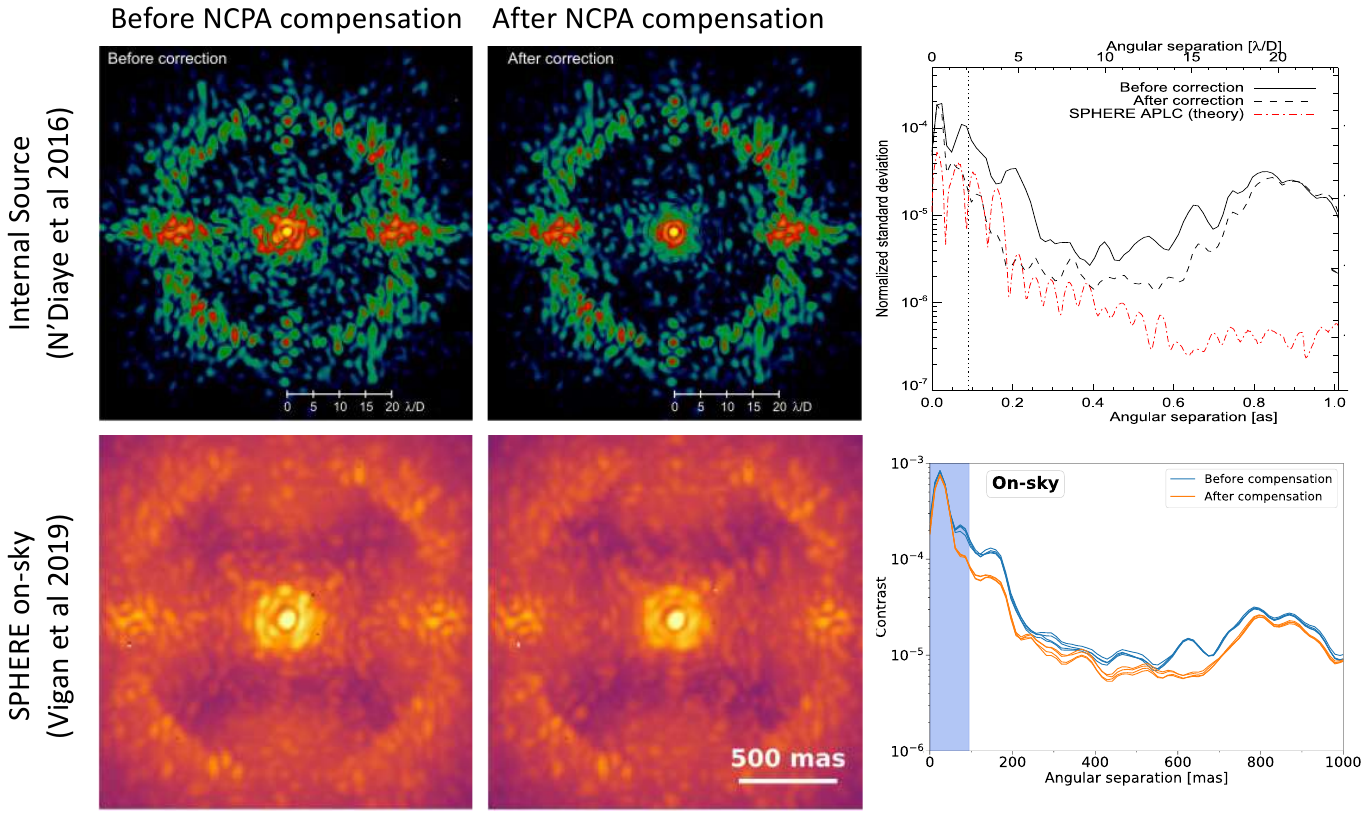}
\end{tabular}
\end{center}
\caption[]{\label{fig:zelda_vigan_sphere} \textbf{Zernike WFS NCPA correction on SPHERE}: internal source (top\cite{ndiaye2016_CalibrationQuasistaticAberrations}), and on-sky (bottom\cite{vigan2019_CalibrationQuasistaticAberrations}). From left to right: before compensation, after compensation, and contrast curves comparison (azimuthal standard deviation profiles of the image normalized to the off-axis PSF). Figure adapted from N'diaye et al. (2016)\cite{ndiaye2016_CalibrationQuasistaticAberrations} and Vigan et al. (2019)\cite{vigan2019_CalibrationQuasistaticAberrations}.}
\end{figure}

At the time of SPHERE's commissioning in 2014, a Zernike WFS \cite{zernike1934_DiffractionTheoryKnifeedge} was installed in the SPHERE instrument, called Zelda (which stands for Zernike sensor for Extremely Low-level Differential Aberrations). The Zernike mask is purposefully located in the coronagraphic focal plane mask wheel of the IRDIS. To measure aberrations, the system is set up in pupil-imaging mode to perform the phase aberration measurements. This configuration allows Zelda to measure the upstream aberrations responsible for the contrast degradation. This technique has been tested successfully on SPHERE internal source \cite{ndiaye2016_CalibrationQuasistaticAberrations} and on-sky \cite{vigan2019_CalibrationQuasistaticAberrations} (see compiled results on Figure~\ref{fig:zelda_vigan_sphere}). Due to problem of loop stability currently under investigation, this Zernike WFS was never completely integrated in SPHERE's standard calibration process, despite showing a clear improvement in contrast performance compared to existing SH reference slopes. 

SH reference slopes were initially set on the ones measured with the phase diversity during commissioning (see Section \ref{subsec:init_sauvagecorrec}). However, a daily NCPA calibration was performed to update them using the Zernike WFS for a period of time. However, this correction was not always stable (slopes were sometimes hitting the limit range of the WFS, creating instabilities or putting the DM in saturation). This daily Zernike measurement is still performed for NCPA monitoring purposes, but not used anymore to modify the SH references slopes. Currently, the SH references slopes are still the ones measured during commissioning. 

The main advantage of this technique is its efficiency in the measurement. Not only the Zernike offers an optimal sensitivity to photon noise at all spatial frequencies \cite{guyon2005_LimitsAdaptiveOptics}, it allows an optical path difference (OPD) map in only a couple of images (on and off the Zernike mask). It is also relatively simple to operate (now that the mask is in the wheel) and has been extremely useful to measure in details the origin and evolution of the SPHERE NCPA (see Section~\ref{sec:measurement_evol})\cite{vigan2022_CalibrationQuasistaticAberrations}.

The main limitation of this technique is that it only allows the calibration of the NCPA and not the minimization of the speckles directly in the focal plane. Figure~\ref{fig:zelda_vigan_sphere} shows that although performance in the focal plane were greatly improved, some speckles remain. These are likely due to amplitude aberrations, that cannot be corrected with this technique. Finally, one can also assume that the replacement of the coronagraph mask by the Zernike mask, even in the same optical plane, probably slightly modify the incoming electrical field.

\subsection{Dark-Hole method}
\label{subsec:darkhole}
\begin{figure} [ht]
\begin{center}
\begin{tabular}{c  c} 
\includegraphics[width=0.75\linewidth]{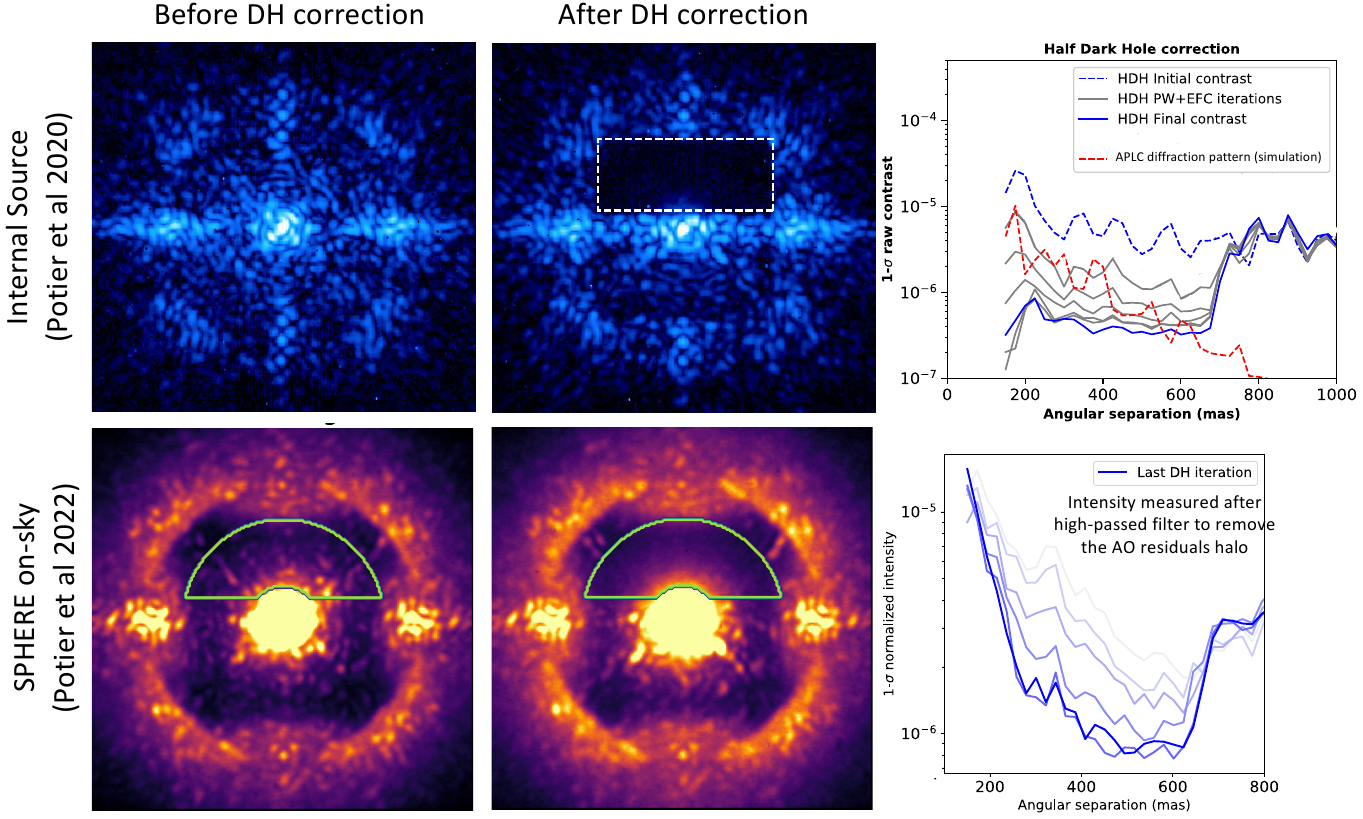}
\end{tabular}
\end{center}
\caption[]{\label{fig:pwp_potier_sphere} \textbf{PWP + DH correction on SPHERE}: internal source (top), and on-sky (bottom). From left to right: before correction, after correction, and contrast curves comparison (azimuthal standard deviation profiles of the image normalized to the off-axis PSF). Figure adapted from Potier et al. (2020, 2022)\cite{potier2020_IncreasingRawContrast, potier2022_AandA_IncreasingRawContrast}.}
\end{figure}

Pair-Wise Probing (PWP) \cite{borde2006_HighContrastImagingSpace,giveon2007_ClosedLoopDM} is a variation of the phase-diversity method for coronagraphic instrument. Probes are known phases introduced by the deformable mirror (usually a single actuator push or a cardinal sinus) to change the focal plane speckles and use a linear model to recover their electrical field. PWP is usually combined with a linear correction algorithm called Electrical Field Conjugation (EFC) which links DM movements to electrical field modification. This correction algorithm goal is to minimize the focal plan energy in a specific zone, called Dark-Hole (DH). 
This technique has been tested successfully on SPHERE internal source \cite{potier2020_IncreasingRawContrast} and on-sky \cite{potier2022_AandA_IncreasingRawContrast} (see compiled results on Figure~\ref{fig:pwp_potier_sphere}).

The main advantage of DH techniques is that the reward they are trying to minimize is not the level of NCPA in the incoming wavefront (minimizing phase), but the starlight energy in the science focal plane (digging a dark-hole). Even for phase-only aberrations at monochromatic light, this is not completely equivalent for a DM with a fixed number of actuators (it is often more interesting to correct only in a specific zones in the focal plane \cite{mazoyer2013_EstimationCorrectionWavefront} to achieve better contrast). But SPHERE speckles have also many origins, with amplitude aberrations (estimated at 8\% RMS for SPHERE\cite{potier2020_IncreasingRawContrast}), chromatic phase aberrations, and diffraction created by the VLT complex apertures and the SPHERE coronagraph. All these sources of speckles are indifferently measured by PWP and can be corrected fully in a half-DH, as shown by the complete removal of speckles after correction with PWP+DH in Figure~\ref{fig:pwp_potier_sphere} (images in central column), compared to the result obtained after calibration with the phase minimization shown in Figures~\ref{fig:coffee_sphere_paul14}~and~\ref{fig:zelda_vigan_sphere} (images in central column). One can also notice the performance of the 2 methods with respect to the level of the simulated coronagraph diffraction in Figures~\ref{fig:zelda_vigan_sphere}~and~\ref{fig:pwp_potier_sphere} (red curves on top right plots). The Zernike WFS is a phase correction method and is theoretically limited to the coronagraph diffraction limit whereas the PWP technique in half Dark-Hole can correct below this level.

There are two main drawbacks of this technique. First, to introduce a probe using SAXO+ DM, it is necessary to modify the reference responses of the PyWFS to push a single actuator by several tens of nanometer (peak-to-valley). As discussed previously for COFFEE, introducing offsets on the PyWFS reference responses is not as straightforward as for SH WFS, and it is currently unclear if the PyWFS will be able to introduce these probes at with which precision.

The second drawback is the time of convergence. To measure precisely the electrical field created by the NCPA, it if necessary to “average the AO residuals”. These aberrations create much faster speckles that cannot be corrected by NCPA techniques, but that are introducing incoherent noise in the “probed” image \cite{singh2019_ActiveMinimizationNoncommon}. By integrating during longer exposures, the AO residual speckles average over a smooth halo that can be disentangled from the NCPA speckles. This is a problem for all NCPA correction techniques on-sky, but the fact that PWP requires more images makes each iteration much longer, using precious telescope time. For the correction obtained in Potier et al. (2022)\cite{potier2022_AandA_IncreasingRawContrast} (Figure~\ref{fig:pwp_potier_sphere}), each probe image acquisition required 80 s (64 s exposure and 16 s overhead, including readout time) for a total of $\sim 400$ s per iteration (four images required for PWP and one image to measure current raw contrast in an “unprobed” image). The minimum time necessary to integrate is dependent on the level of the introduced probe, the level of NCPA (one could imagine variable integration time over the iterations, with shorter probes at the beginning of the correction, increasing as the NCPA level decrease), the amount of AO residuals and their relative dynamic of the NCPA speckles and AO residuals. For SAXO+, one can assume that with a much faster second AO stage frequency (faster and lower AO residuals) and an expected slower evolution of the NCPA (discussed in Section~\ref{sec:measurement_evol}), the integration time could be shorter. These aspects will be studied in simulation in the coming months.

\section{Analysis of IRDIS NCPAs and their temporal evolution in the context of SAXO+}
\label{sec:measurement_evol}

Vigan et al. (2022)\cite{vigan2022_CalibrationQuasistaticAberrations} used the Zernike WFS to perform a deep characterization of the current SPHERE NCPAs level and their temporal stability. We report here its main findings in the context of the SAXO+ upgrade. 

\subsection{Current NCPA level and frequency content}
\begin{figure} [ht]
\begin{center}
\includegraphics[width=\linewidth]{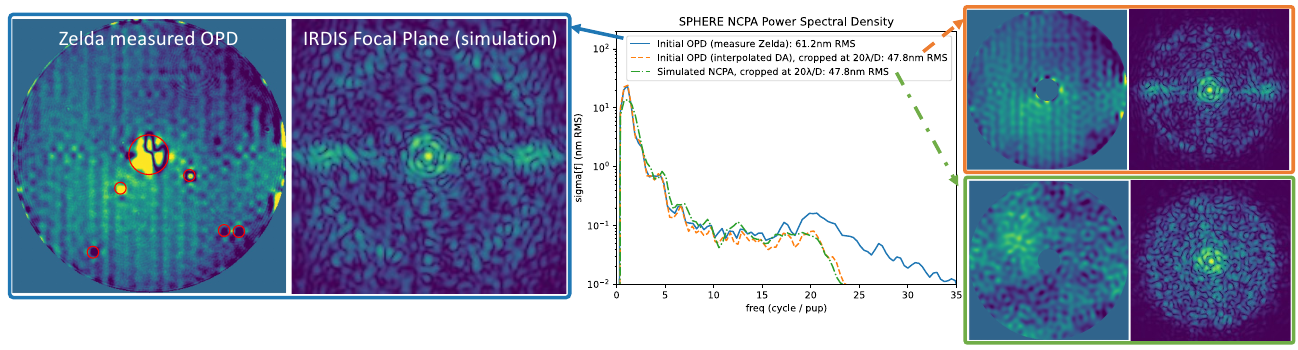}
\end{center}
\caption[]{\label{fig:sphere_ncpa_psd} \textbf{SPHERE measured NCPAs.} Left: Zelda measured OPD map and associated IRDIS coronagraphic image. We circled on the left the dead actuators that are masked and interpolated and the part behind VLT aperture's central obscuration that is hidden. Center: PSD of the NCPAs. Top, Right : Zelda OPD map where dead actuators were interpolated and where the frequencies after the SAXO cut-off are removed and associated IRDIS coronagraphic image. Bottom, Right: simulated opd map with the same PSD and the same RMS level and associated IRDIS coronagraphic image.}
\end{figure} 
An example of an NCPA OPD map measured with Zelda in internal source is presented in Figure~\ref{fig:sphere_ncpa_psd} (left) as well as its measured Power Spectral Density (PSD). We first notice a lot of aberrations in the central part of the pupil (indicated by a red circle), because this zone of the pupil is behind the VLT pupil / SPHERE apodizer central obscurations and DM actuators in this position are not controlled. We also clearly notice the dead actuators on the first DM (also circled in red). These should normally be seen by the pyramid and corrected by the SAXO+ DM, although it might require a significant amount of stroke with the SAXO+ DM. This important question is out of the scope of this paper, as we focused on SAXO+ NCPA. After masking these artifacts, the level of aberrations is typically between 55 and 60 nm RMS, including  45-50 nm RMS in the SAXO correction zone ($<20\lambda/D$) and 40-45 nm RMS in the SAXO+ correction zone ($<13 \lambda/D$).

This map has been obtained in internal source and when the SH reference slopes are set to the default one measured during commissioning. Two bright patterns in the horizontal direction are also very clear in the associated simulated IRDIS image and are present in all SPHERE/IRDIS scientific images (see on-sky images in Figures~\ref{fig:zelda_vigan_sphere}~and~\ref{fig:pwp_potier_sphere}). These are due to characteristic frequencies present along the principal direction of the DM (vertical stripes). At least part of these frequencies are within the cut-off of the first stage DM (creating speckles inside the correction ring) and should be corrected by recalibrating the SH reference slopes. The correction of these speckles are actually very clear with Zelda, with some of the best gain in contrast appears to be around 650 mas (Figure~\ref{fig:zelda_vigan_sphere}, right).

Apart from these features (dead actuator and uncorrected NCPA from the first stage), as new optics will be introduced in the beam path after the second stage WFS, we are currently assuming that the level of aberrations and frequency content after SAXO+ will be comparable to the ones presented in this section.

\subsection{NCPA temporal evolution}

\begin{figure} [ht]
\begin{center}
\begin{tabular}{c  c} 
\includegraphics[width=0.7\linewidth]{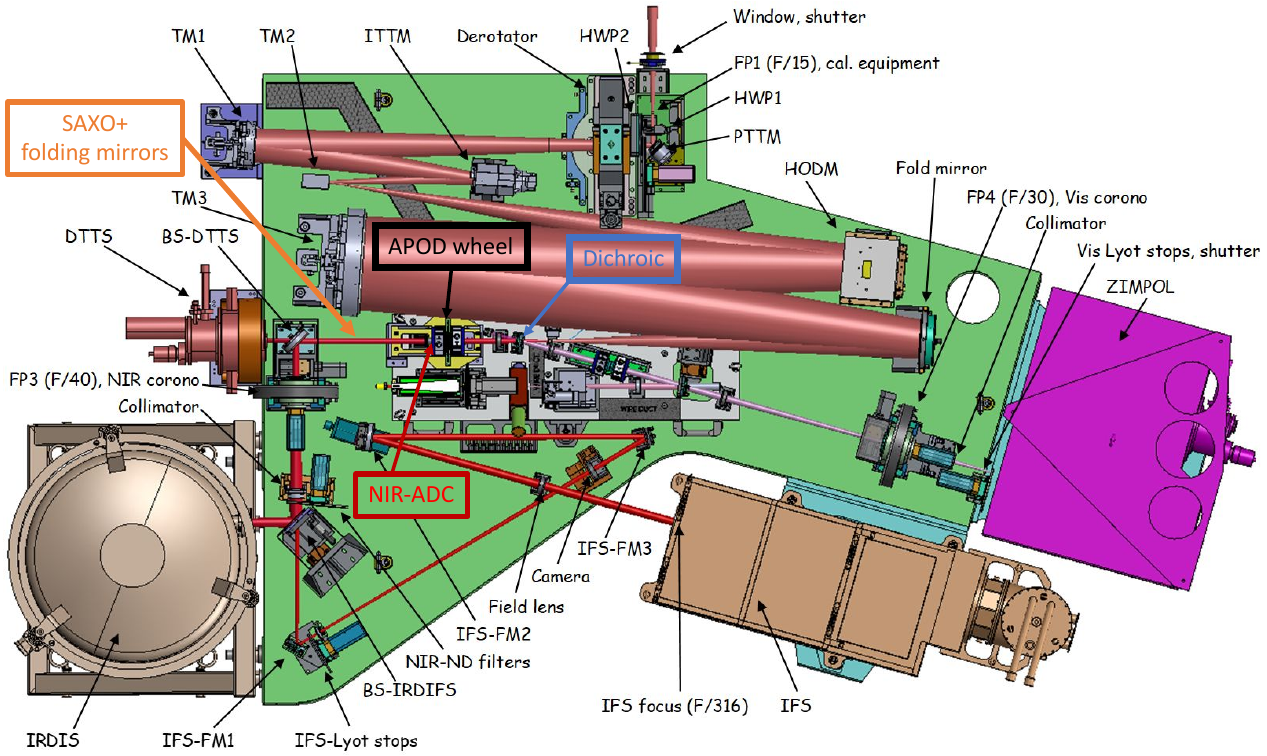}
\end{tabular}
\end{center}
\caption[]{\label{fig:sphere_layout} \textbf{SPHERE Layout}. Boxed labels indicate important elements for NCPAs: the dichroic sending the visible light toward the SAXO WFS, the apodizer wheel and right after, the NIR ADC. The SAXO+ pick up mirrors sending light to Y/J light to the second stage WFS will be installed between the ADC and the DTTS\cite{stadler2022_SAXOSecondstageAdaptive}. Figure adapted from Beuzit et al. (2019)\cite{beuzit2019_SPHEREExoplanetImager}.}
\end{figure}

Soon after the commissioning, a study of the speckle correlation time was carried out on SPHERE, showing a fast decorrelation with an exponential decrease for timescales under a second.\cite{milli2016_SpeckleLifetimeXAO}. The origin of this fast internal turbulence was investigated using Zelda\cite{vigan2022_CalibrationQuasistaticAberrations}. The first source of fast turbulence investigated was the airflow pumped into the SPHERE's enclosure to maintain a clean environment. However, tests conducted with and without the airflow showed no difference in the amplitude of the internal turbulence. The second suspected source was the motor of the apodizer wheel, located in a pupil plane close to the NIR atmospheric dispersion corrector\cite{vigan2022_CalibrationQuasistaticAberrations} (see Figure~\ref{fig:sphere_layout}). This motor is located underneath the optical beam, downstream of the visible-NIR dichroic filter sending the light to SAXO WFS. According to Vigan et al. (2022) \cite{vigan2022_CalibrationQuasistaticAberrations}, it is “extremely likely” that this motor is responsible for the observed fast turbulence due to the high temperature gradient it introduces. Fortunately, the SAXO+ pick-up mirrors will be installed after the NIR atmospheric dispersion correctors (ADC) and the differential tip-tilt sensor (DTTS)\cite{stadler2022_SAXOSecondstageAdaptive} as shown on Figure~\ref{fig:sphere_layout}). This turbulence will therefore be measured and corrected by the PyWFS. It is also the case for the NIR ADC that was also identified by these authors as a likely culprit for part of the measured NCPA evolution. 
The general level of NCPA will probably be equivalent in SAXO+ to what it is currently in SAXO. However, two of the main sources of the evolution of NCPA in the current system (heat from apodizer electronics accounting for fast speckle decorrelation and NIR ADC probably responsible for part of the slower NCPA evolution) will now be located in the common path of the second system. For this reason, we can probably expect more stable NCPA in SAXO+. Simulation of hour-long sequences with evolving NCPA will be done in the coming months to create realistic simulation to evaluate performance after post-processing. 

\section{NCPA CORRECTION WITH SAXO+: SIMULATIONS}
\label{sec:ncpa_corr}

In this section, we show the result of simulations to test the correction of NCPAs in the case of perfect estimation. All simulations in this section used \texttt{COMPASS}\cite{gratadour2016_COMPASSStatusUpdate}, an end-to-end simulation tool already used to design other AO systems. We chose 4 given star/observation cases (a bright star with 2 different observing conditions -best and worst- and 2 fainter stars with medium conditions) with AO parameters optimized by SAXO+'s AO team. We use SAXO+'s reference controller, defined in more details in Béchet et al. (2023) \cite{bechet2023_InverseProblemApproach} and its parameters were optimized to return best contrast. This controller assumes a stand-alone case where the two loops are independent, meaning one real time computer (RTC) for each loop. Other controllers\cite{bechet2023_InverseProblemApproach, galland2023_DisentangledCascadeAdaptive} are also studied as part of SAXO+ upgrade, as well as the optimization of the PyWFS parameters\cite{goulas2024numericalsimulationssaxoupgrade} (central wavelength, modulation radius, frequency). In this paper, we only use one set of parameters to focus on the impact of NCPAs. Table~\ref{tab:simu_param} describes the parameters used in the simulations. 

\begin{figure} [ht]
\begin{center}
\begin{tabular}{c c} 
\includegraphics[width=0.45\linewidth]{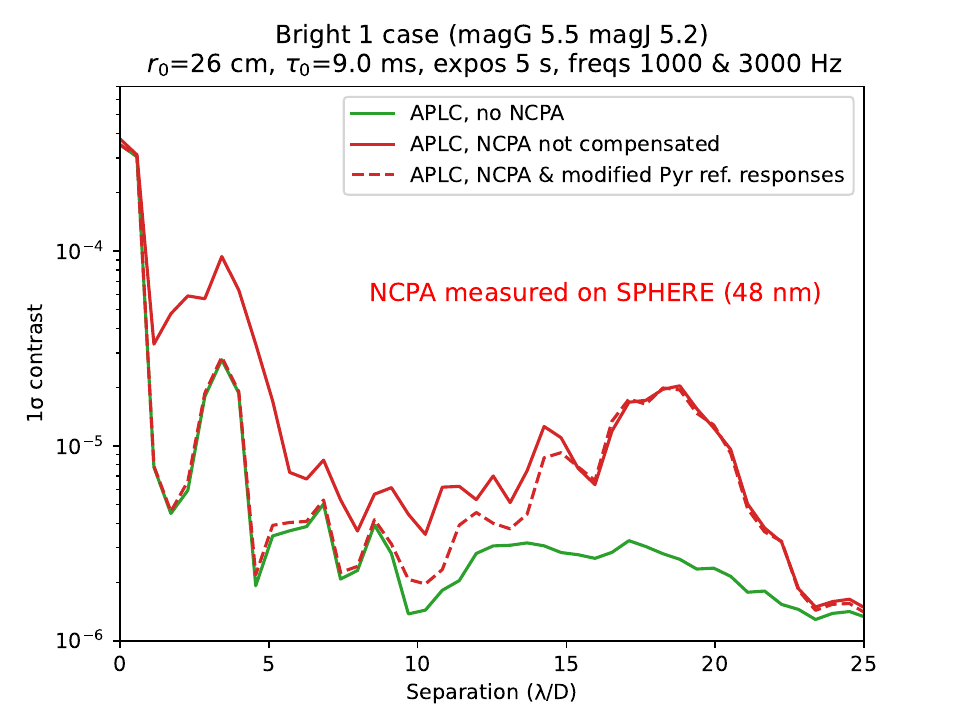}
&
\includegraphics[width=0.45\linewidth]{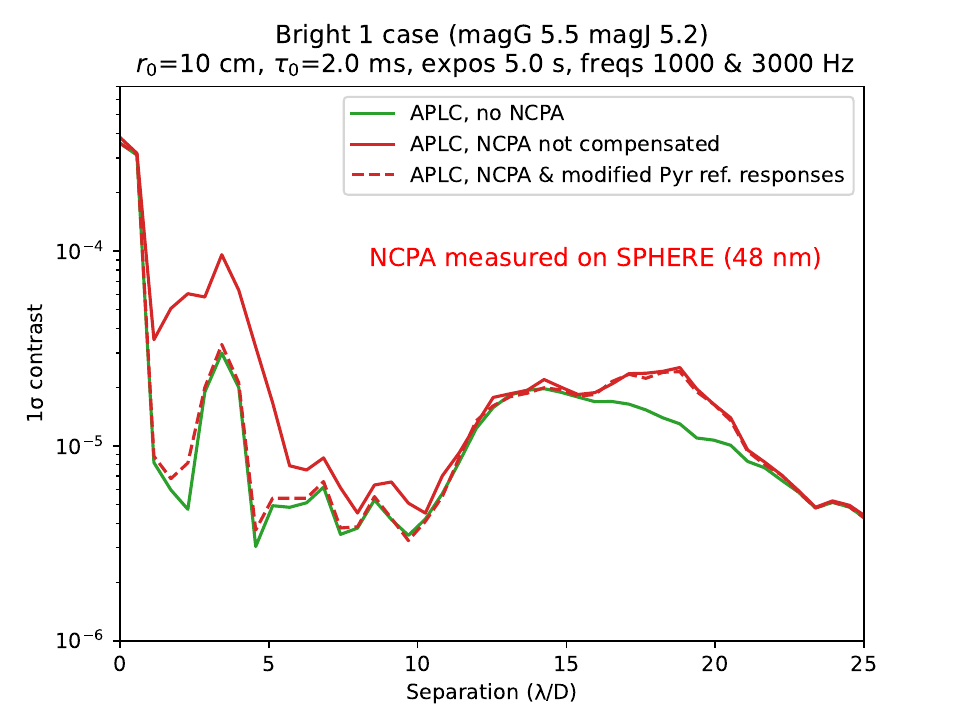}\\
\includegraphics[width=0.45\linewidth]{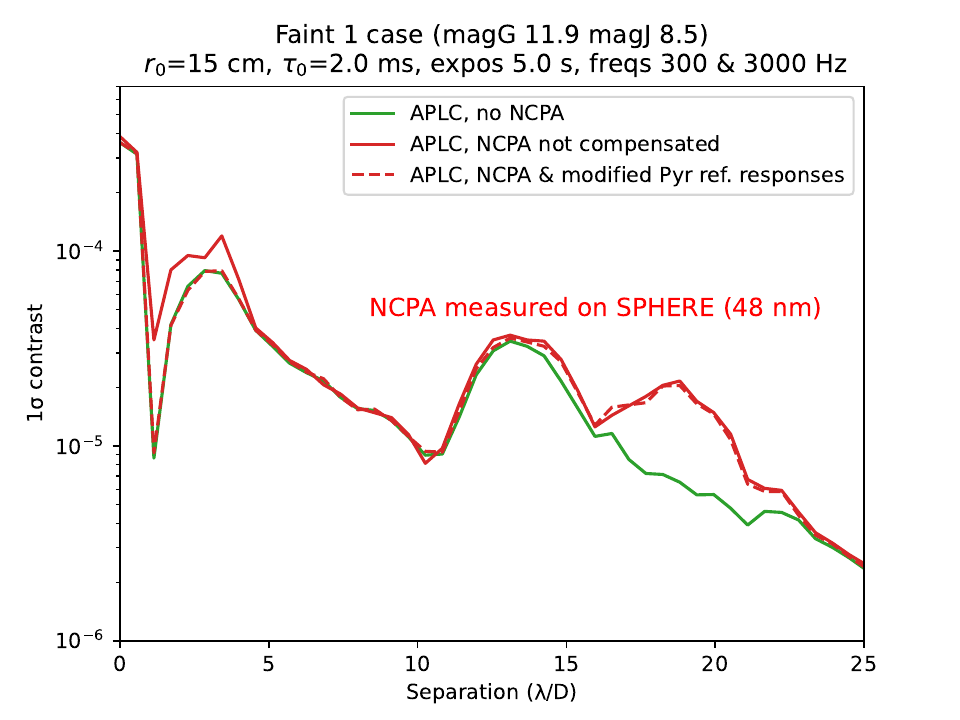}
&
\includegraphics[width=0.45\linewidth]{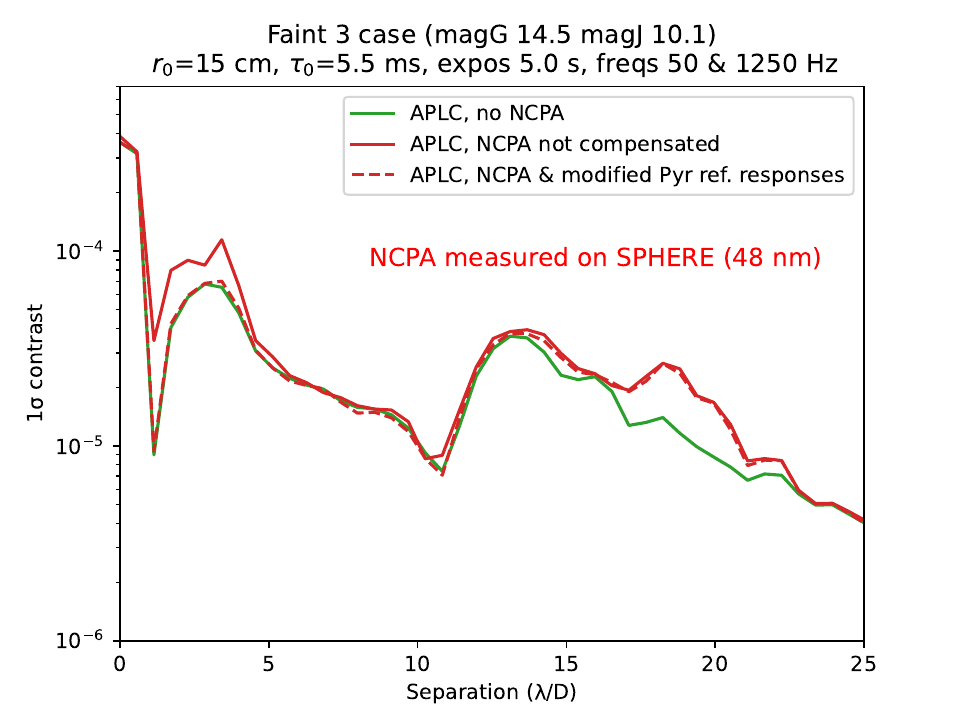}
\end{tabular}
\end{center}
\caption[]{\label{fig:realncpacase}\textbf{Simulated IRDIS performance in presence of NCPAs}. Contrast curves (azimuthal standard deviation profiles of the image, normalized to the off-axis PSF, as a function of the distance to the star) without NCPAs (green solid line), with non-compensated NCPA (red solid line) and after compensation by pyramid reference responses modification (red dashed line). Simulation conducted with an APLC and the SPHERE measured NCPAs (48 nm RMS) in the 4 identified cases.}
\end{figure} 

We used 2 types of coronagraphs, a perfect coronagraph \cite{cavarroc2006_FundamentalLimitationsEarthlike} and the current SPHERE Apodized Lyot Stop Coronagraph (APLC \cite{boccaletti2008_PrototypingCoronagraphsExoplanet}). The former can measure the “ideal” performance of the AO system without limitations from the coronagraph diffraction pattern. The latter is used to simulate a more realistic response of the SAXO+ instrument to a given level of NCPA. In this study, we only study the response of SAXO+ to NCPA, as the performance and correction of NCPA with SAXO have already been well studied on SPHERE. 

NCPA OPD maps are introduced after SAXO+ second loop. We did not introduce aberrations between the two WFS, assuming that those will be corrected by the second loop for SAXO+ accessible frequencies. The NCPA OPD map introduced were based on the measurement done on SPHERE using the Zernike WFS (shown in Figure~\ref{fig:sphere_ncpa_psd}). We first introduced the real OPD map where we removed the dead actuators by interpolating, and filtered all frequency above the SAXO's cut-off frequency (20 $\lambda/D$), as the aberrations measured by the Zernike WFS beyond this frequency are in majority AO residuals already included in the simulation. This OPD map is presented in Figure~\ref{fig:sphere_ncpa_psd} (top, right). 

Figure~\ref{fig:realncpacase} shows the results in contrast for the four simulated cases with an APLC coronagraph. We first ran a long pause image (5s)simulation  without any NCPA showing the performance only limited by the atmospheric turbulence and/or coronagraph diffraction (green curves). We then introduced NCPAs in the coronagraphic channel and ran the same simulation without trying to compensate for NCPAs, showing a degradation of the performance in contrast (red curves). Finally, we compensated for the NCPA assuming a perfect estimation. Once the pyWFS controller matrices were measured during the simulation initialization in \texttt{COMPASS}, the same OPD map was introduced both in the SAXO+ PyWFS channel and the coronagraph channel. The PyWFS responses were measured and then used to update the pyWFS reference responses. Finally, we removed the NCPA OPD map in the SAXO+ PyWFS channel but left it in the coronagraph channel and ran the simulation (red dotted curves). No optimization of the optical gains were performed in these preliminary tests.

From this figure, we can conclude:
\begin{itemize}
    \item in the case of the bright stars, both in good (Bright1 best) and bad (Bright1 worse) atmospheric conditions, the performance of SAXO+ will be heavily limited by the level of NCPA currently measured on SPHERE. 
    \item in the fainter cases, the amount of NCPA normally present on the SPHERE as measured by the Zernike Wavefront sensor will barely impact the performance, as most of the coronagraphic focal plan will be limited by AO residuals. 
    \item finally, in all cases, the modification of the pyWFS reference responses can calibrate the NCPA and return to the expected contrast level predicted for these observing conditions in the absence of NCPA. 
\end{itemize}

\begin{figure} [ht]
\begin{center}
\begin{tabular}{c  c} 
\includegraphics[width=0.48\linewidth]{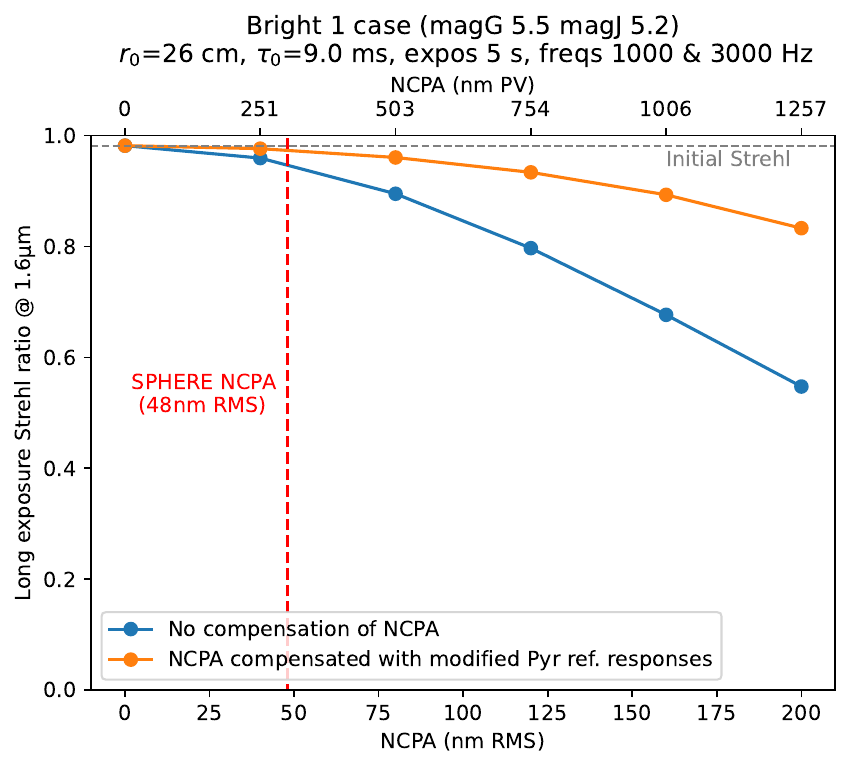}
&
\includegraphics[width=0.48
\linewidth]{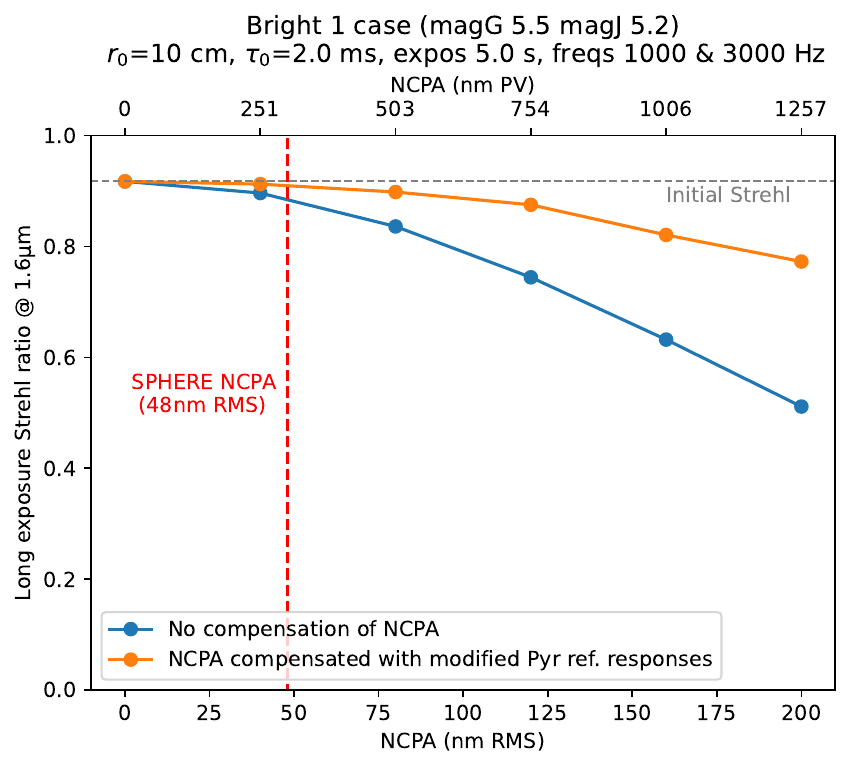}
\end{tabular}
\end{center}
\caption[]{\label{fig:increasing_ncpa_SR}\textbf{Strehl ratio vs NCPA level} (same PSD as the SPHERE NCPA with increasing level) for a bright star and 2 different atmospheric conditions.}
\end{figure} 

\begin{figure} [ht]
\begin{center}
\begin{tabular}{c c} 
\includegraphics[width=0.48\linewidth]{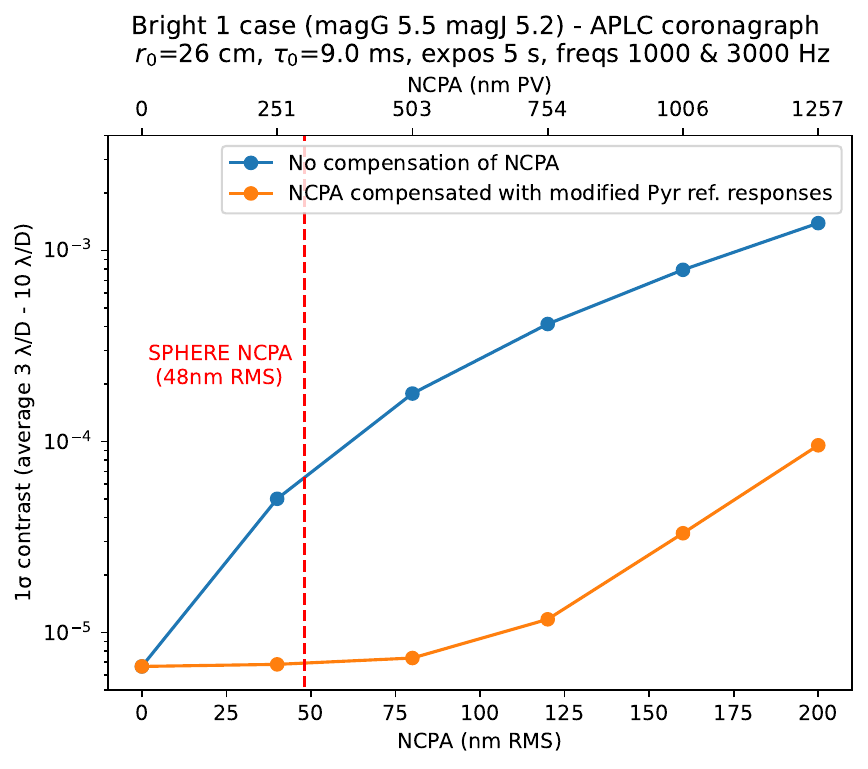}
&
\includegraphics[width=0.48
\linewidth]{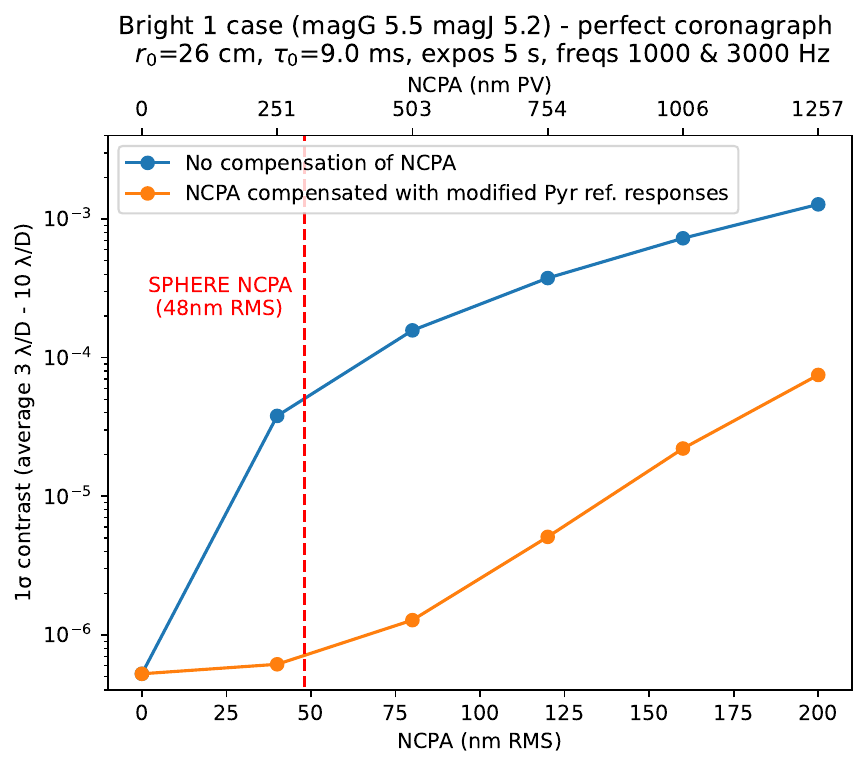}\\
\includegraphics[width=0.48\linewidth]{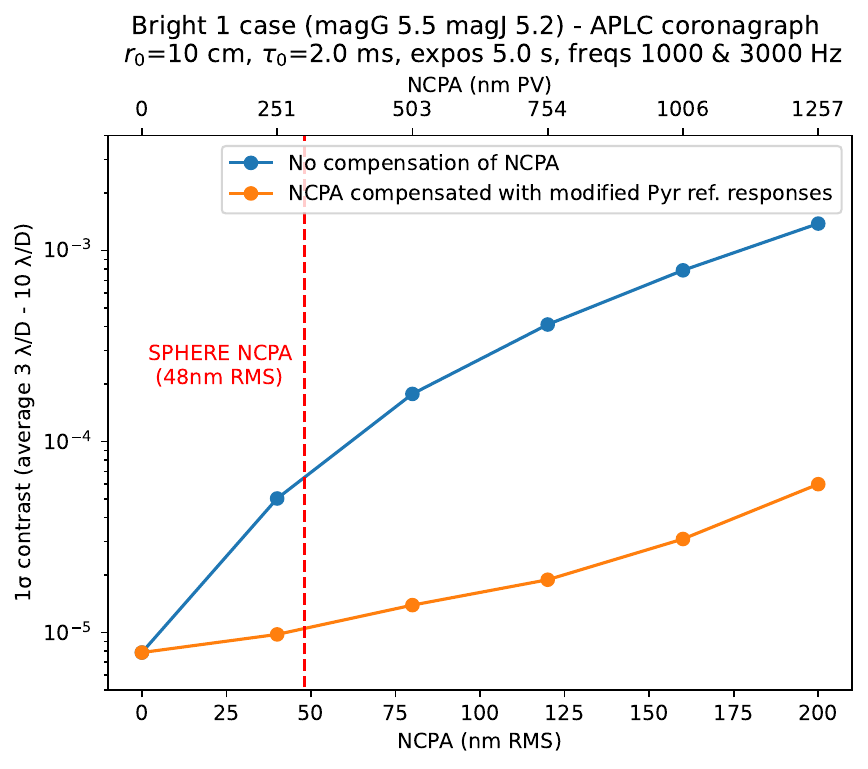}
&
\includegraphics[width=0.48
\linewidth]{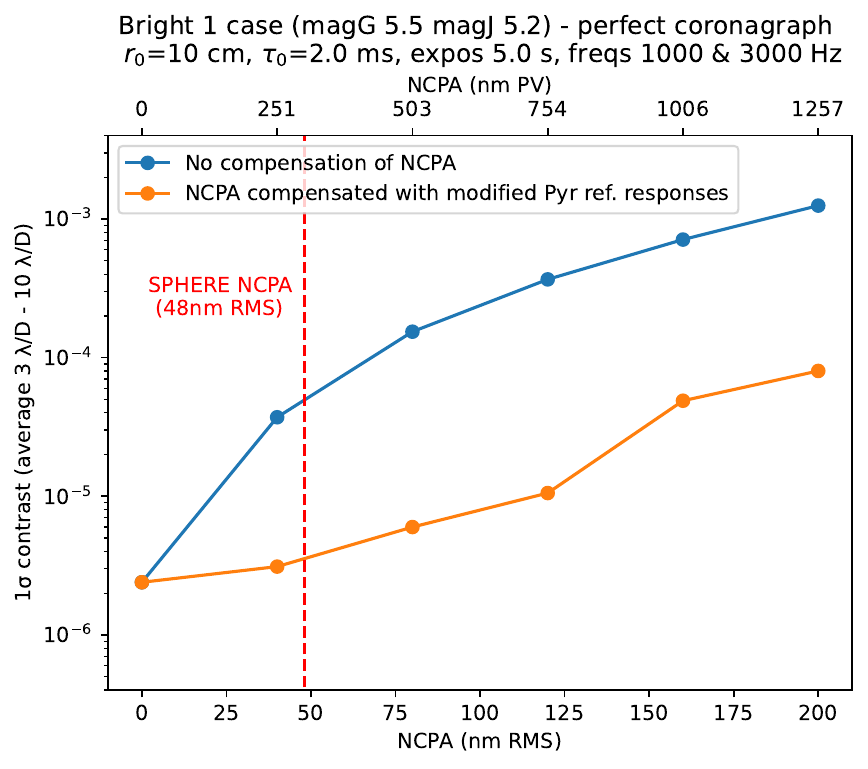}
\end{tabular}
\end{center}
\caption[]{\label{fig:increasing_ncpa_cont} \textbf{Mean contrast vs NCPA level} for a bright star in 2 different atmospheric conditions for 2 different coronagraphs (APLC and perfect coronagraph). Mean contrast is the average between 3 and 10 $\lambda/D$ of the azimuthal standard deviation profiles of the image, normalized to the off-axis PSF) }
\end{figure}

In a second part, we then simulated a random OPD map with the same PSD and same level of aberration at the latter (48 nm RMS in the VLT pupil). This allows us to make the OPD more uniform azimuthally and to remove frequencies in a specific direction that are specific to the first stage DM (this OPD is shown in Figure~\ref{fig:sphere_ncpa_psd} bottom right). This simulated OPD is then scaled at different levels (from 40 nm RMS to 200 nm RMS in the VLT pupil) to measure the capacity of the system to correct for these increasing NCPA amplitude level. The result is shown first on the impact on Strehl Ratio (at IRDIS wavelength, 1.6$\mu$m) in Figure~\ref{fig:increasing_ncpa_SR} and on the final contrast in Figure~\ref{fig:increasing_ncpa_cont}, only for the brightest cases. We indicated using a red vertical line at 48 nm RMS the level of aberrations currently measured on SAXO in the SAXO correction zone ($<20 \lambda/D$).

Both in good and bad observing conditions, the modification of the PyWFS reference responses allows the Strehl Ratio to return to performance in the absence of NCPA up to 80 nm of NCPA, corresponding to $\sim$500 nm peak-to-valley (PV) with the given PSD. Above this amplitude, the pyramid cannot fully correct for those NCPAs and they impact significantly the Strehl Ratio. 

This effect is also visible in the contrast level, more sensitive  to small errors in the wavefront at all frequency than the Strehl Ratio. The range of allowed NCPA before significant contrast degradation is larger in the case of the APLC (Figure~\ref{fig:increasing_ncpa_cont}, left plots) than for the perfect coronagraph (Figure~\ref{fig:increasing_ncpa_cont}, right plots), which is expected as the APLC base contrast level is often limited by the coronagraph diffraction and also because this coronagraph is much less sensitive to low-order aberrations. Finally, the range of allowed NCPA is larger in the best observing conditions (top) than in the worst observing conditions (bottom) which is also expected. 

In all cases, we deduce from these preliminary tests that with the current level of NCPA ($\sim$ 50 nm RMS) measured on SPHERE using the Zernike WFS, we expect that we can correct them using the modification of the PyWFS reference responses. This promising result can most likely be explained by the fact that SAXO+ pyramid is a second stage and for this reason, the level of Strehl Ratio seen by this WFS is much higher than in a lot of future AO system with a pyramid as a single WFS. Another reason is that SAXO+ PyWFS is in Y/J and is less sensitive to same OPD offsets than other instruments using pyramid in visible. With the SPHERE (already operational) Zernike WFS (which does not require a DM-created diversity and will be unaffected by the second stage pyramid) we can confidentially predict that we will be able to reduce NCPA levels below the existing level on SPHERE in most conditions. We will continue our analysis to assess the feasibility of the other WFS techniques, which require the introduction of probes on the PyWFS reference responses. 

\section{CONCLUSION AND FUTURE WORK}

The constant development of several NCPA estimation / correction methods (COFFEE, ZELDA, PWP+DH method) using SPHERE internal source and ESO's calibration programs in the decade since SPHERE commissioning has allowed them to increase their readiness to be now included in the SAXO+ toolbox. The goal is to test these methods of NCPA estimation / correction with SAXO+ to understand their advantages and defects in the context of PCS. 

However, the different nature of the SAXO+ WFS (a pyramid instead of a SH WFS) raise new challenges in the application of the these methods. Fortunately, SAXO+ PyWFS is a second stage and in Y/J and is therefore much less sensitive to same OPD offsets than other instruments (single PyWFS and/or in the visible). It appears that, at the level of NCPA currently measured on SPHERE, the pyramid will be able to mostly compensate for NCPAs in order to increase the contrast levels to the limit allowed by the AO loop. Additional simulations are necessary to evaluate if a technique based on DM-introduced diversity (COFFEE, PWP+DH) will be achievable, and with which precision. 

More simulations are also necessary to understand in all conditions observed in Paranal, and for all star brightnesses, and derive the exact requirements in NCPA correction level so that they do not limit the performance in contrast. Most of the “after compensation” simulations shown in this paper are limited by the diffraction of SPHERE APLC. Coronagraph performance (in contrast and throughput) in the presence of complex apertures and central obscuration (e.g. in the case of APLC \cite{ndiaye2015_ApodizedPupilLyot, ndiaye2016_ApodizedPupilLyot}). An upgrade of the apodizers will be performed as part of SPHERE+ and we will perform simulation with new apodizers. 

Finally, as described in Section~\ref{sec:measurement_evol}, the position of the second loop in the SPHERE instrument is ideal as it brings back in the common path two major sources of turbulence / evolution of the NCPA identified in the current instrument: the apodizer wheel electronics heating the air underneath the beam and the NIR ADC. SAXO+ will therefore encounter much more stable NCPAs which will increase the performance of NCPA active correction techniques described in this paper, as well as post-processing methods that will be used on the science images.

\noindent \textit{Software} - This study used the following python packages: \texttt{Asterix}\cite{mazoyer_AsterixSimulator}, 
\texttt{Astropy}\cite{astropy2022},
\texttt{COMPASS}\cite{gratadour2016_COMPASSStatusUpdate} and 
\texttt{pyZELDA}\cite{Vigan18_pyZELDA}.

\section*{APPENDIX: PARAMETERS USED IN \texttt{COMPASS} SIMULATIONS}
\begin{table}[!ht]
\caption{\label{tab:simu_param} \texttt{COMPASS} simulation parameters used for the 4 cases studied}
    \centering
    \small
    \begin{tabular}{l c c c c }
        \textbf{Parameters} & \textbf{Bright 1 best} & \textbf{Bright 1 worse} & \textbf{Faint 1 fast} & \textbf{Faint 3 medium} \\ \hline \hline
         ~ \\ 
        \multicolumn{5}{l}{\textbf{Star parameters}} \\ \hline
        Mag G & 5.5 & 5.5 & 11.9 & 14.5 \\ 
        Mag J & 5.2 & 5.2 & 8.5 & 10.1 \\ \hline \hline
        ~ \\ 
        \multicolumn{5}{l}{\textbf{SAXO+ system}} \\ \hline
        Stage 1 DM & \multicolumn{4}{c}{41x41 + TT} \\ 
        Stage 1 WFS & \multicolumn{4}{c}{SH WFS (40x40 subapertures)}  \\ 
        Stage1 central wavelength [$\mu${m}] & \multicolumn{4}{c}{0.7}  \\ 
        Stage 1 readout noise [e-/pix] & \multicolumn{4}{c}{0.1} \\ \hline
        Stage 2 DM & \multicolumn{4}{c}{28x28}  \\ 
        Stage 2 WFS & \multicolumn{4}{c}{py WFS (50x50 subapertures)}  \\ 
        PyWFS modulation radius [$\lambda/D$] & \multicolumn{4}{c}{3} \\ 
        Stage 2 central wavelength [$\mu${m}] & 1.043 & 1.043 & 1.043 & 1.144 \\ 
        Stage 2 readout noise [e-/pix] & \multicolumn{4}{c}{0.3}  \\ \hline \hline
        ~ \\ 
        \multicolumn{5}{l}{\textbf{Flux \& turbulence}} \\ \hline
        Stage 1 photFlux [photons/m2/s] & 1.06E+07 & 1.06E+07 & 2.92E+04 & 2.66E+03 \\ 
        Stage2 photFlux [e-/m2/s] & 9.37E+06 & 9.37E+06 & 7.11E+05 & 1.63E+05 \\ 
        Seeing [arcseconds] & 0.4 & 1 & 0.7 & 0.7 \\ 
        Coherence Time [ms] & 9 & 2 & 2 & 5.5 \\ \hline \hline
        ~ \\ 
        \multicolumn{5}{l}{\textbf{Controller parameters}} \\ \hline
        Stage 1 frequency [Hz] & 1000 & 1000 & 300 & 50 \\ 
        Stage 1 gain & 0.2 & 0.4 & 0.05 & 0.05 \\ 
        Stage 1 \# controlled mode & 1200 & 1200 & 400 & 400 \\\hline 
        Stage 2 frequency [Hz] & 3000 & 3000 & 3000 & 1250 \\ 
        Stage 2 gain & 0.2 & 0.5 & 0.3 & 0.3 \\ 
        Stage 2 \# controlled modes & 400 & 540 & 540 & 540 \\ \hline \hline
        ~ \\
        \multicolumn{5}{l}{\textbf{IRDIS parameters}} \\ \hline
        Exposure time [s] & \multicolumn{4}{c}{5} \\
        Imaging wavelength [$\mu${m}] & \multicolumn{4}{c}{1.65} \\ 
    \end{tabular}
    
\end{table}

\bibliography{report} 
\bibliographystyle{spiebib} 

\end{document}